\documentclass{article}
\usepackage[T1]{fontenc}
\usepackage[utf8]{inputenc}
\usepackage{spconf,amsmath,graphicx,hyperref}
\usepackage{times}
\usepackage{latexsym}
\usepackage{multirow}
\usepackage{booktabs}
\usepackage{microtype}
\usepackage{inconsolata}
\usepackage{graphicx}
\usepackage{lipsum}
\usepackage{xcolor}
\usepackage{orcidlink}

\title{Toward Complete Speaker Anonymization: \\a Streaming Cascaded System for Voice Privacy and PII Removal}

\name{Stanisław Kacprzak\,\orcidlink{0000-0002-8717-9327}\,$^{1,2}$\thanks{This research was supported by the Excellence Initiative – Research University Programme
for the AGH University of Krakow and, in part, by the National Science Centre, Poland under Grant 2021/42/E/ST7/00452. For the purpose of Open Access, the author has applied a CC-BY public copyright licence to any Author Accepted Manuscript (AAM) version arising from this submission.} \qquad Thomas Thebaud\,\orcidlink{0000-0001-8953-7872}\,$^{1,3}$ \qquad Najim Dehak\,\orcidlink{0000-0002-4489-5753}\,$^{1}$}
  
  \address{$^{1}$\,Center for Language and Speech Processing, Johns Hopkins University, Baltimore, Maryland, USA \\
        $^{2}$\,AGH University of Krakow, Institute of Electronics, Krakow, Poland \\
      $^{3}$\,AMIAD, Palaiseau, FR \\
       }

\begin{document}
\maketitle
\raggedbottom
\begin{abstract}
We present a cascade streaming voice privacy system, consisting of ASR, NLP, and TTS modules. By leveraging recent improvements in all three aspects of speech and language processing, we show that such a system is capable of achieving a result comparable to state-of-the-art offline voice privacy systems, while providing controllable capabilities of content protection beyond voice. With an estimated utterance-onset end-to-end latency of 717\,ms, the presented system preserves speech content with 11\% WER on Librispeech, while detecting PII with a 69\% SLU F1 score and achieving up to 47.66\% EER against a semi-informed attacker. 
\end{abstract}

\section{Introduction}
Privacy and the protection of personal information have become central in machine learning in recent years.
The General Data Protection Regulation (GDPR)~\cite{gdpr2018general} has set the grounds for strong protection of personal data for all European citizens.
Personal data includes both biometric data (such as speech, face images, or fingerprints) and any \textbf{personal identifying information} (PII), which includes direct identifiers (names, home addresses), digital identifiers (IP addresses), and sensitive information (religious beliefs, sexual orientation).

\textit{Speech anonymization} is the technique of removing one's identity from a speech signal while keeping useful information such as the linguistic or emotional content of the speech.
Driven by initiatives such as the Voice Privacy Challenge~\cite{tomashenko2024voiceprivacy,miao2026voiceprivacy} (VPC), the speech community has shown a variety of techniques allowing the removal of acoustic and prosodic cues in a signal such that an individual can no longer be recognized from their anonymized speech, even by informed systems. 
Currently, neural speech anonymization systems generally follow two main approaches:
Voice Conversion (VC) and cascade systems. VC systems~\cite{walczyna2023overview}, which directly transform the speech, one short temporal window at a time, to modify the speaker's identity to sound like a target identity. Those systems usually offer weaker privacy, but allow for a stronger conservation of utility signals~\cite{panariello2024voiceprivacy}, and can allow streaming processing of speech~\cite{kanagawa2023vc, yang2024streamvc}.
Cascaded systems, on the other hand, consisting of Automatic Speech Recognition (ASR) followed by a Text-To-Speech (TTS) system~\cite{aggazzotti2026content, lee2024voice}, have demonstrated strong privacy performance, at the cost of reduced utility. Another cascade system SecureSpeech consisting of ASR, Name Entity Recognition (NER) and TTS was presented in \cite{hui2025securespeech}. However, these systems are designed to work offline.
In \cite{gouvea2023trustera} the authors show redaction of the PII in live speech by masking the audio with a beep sound. %

However, if some systems use cascaded ASR+TTS to remove almost all identifying information from audio \cite{lee2024voice, xinyuan2024hltcoe}, and some systems use cascaded ASR+NER to redact PII \cite{gouvea2023trustera}, the combination of both approaches has seen little exploration.
We propose to apply an addition NLP module to remove PII, creating a cascaded ASR + NLP + TTS system. 
Moreover, we select the components of our cascaded system to make it as fast as possible, to close the gap toward \textit{streaming} anonymization systems. 

An additional motivation for streaming voice privacy systems is their potential use in privacy-sensitive settings, such as remote therapy or support hotlines, where protecting a speaker’s identity may lower barriers to using these services and facilitate more open communication.
Our contributions can be expressed as follows:
\begin{itemize}
    \item We introduce a speech anonymization system that can detect personal information from audio and text through a cascade of an ASR, an NLP, and a TTS modules.
    \item Our system can be used in an almost-streaming fashion, with fast components, which sets up the baseline for fully anonymized streaming systems.
    \item The evaluation of the proposed system. %
\end{itemize}

\section{System components}
\subsection{Automatic Speech Recognition (ASR)}
Since our goal is real-time natural speech generation, processing phrases instead of single words seems to be a better suited for downstream TTS. For that reason, after preliminary experiments with streaming ASR systems, we decided to continue with Whisper \cite{whisper}. Whisper is not an intrinsically streaming ASR; however, it has been shown to work in a streaming scenario \cite{machavcek2023turning}. 
We implement our own algorithm to use Whisper \emph{turbo} and SileroVAD \cite{silerovad} in a streaming manner. We start with 700\,ms audio chunks and process only blocks that contain speech. If less than 100\,ms of non-speech remains after the detected end of speech in a block, blocks of up to 200\,ms are added and VAD processing is repeated until at least 100
\,ms of trailing non-speech is found or the recognition block reaches 1.3\,s. If VAD detects further speech after that internal non-speech segment of at least 100\,ms, the audio corresponding to the next utterance is carried over to the next recognition block. ASR recognition is performed once on the established recognition block. If speech extends to within 100\,ms of the maximum 1.3\,s recognition block boundary, words that may be truncated by the boundary are removed from the transcript, and the corresponding audio is carried over to the next recognition block.

\subsection{PII detection with LLM/NER}
For the detection of personal identifying information (PII), we consider using LLMs and dedicated name entity recognition (NER) models. We choose to test two Qwen2.5-Instruct \cite{qwen2.5} models \cite{qwen_small,qwen_bigger}, two GLiNER \cite{gliner} based models \cite{nvidia_gliner_pii, gliner_community} and \emph{en\_core\_web\_trf} pipeline from spaCy \cite{spacy}.

\subsection{Text-to-Speech (TTS)}
For speech synthesis, we use VoxCPM 2.0 \cite{voxcpm2} with the Nano-vLLM-VoxCPM
inference engine \cite{nano-llm}. This selection was motivated by the ability to perform high-fidelity continuation cloning and native streaming support. Continuation cloning is an extremely desired feature, since our system operates on small chunks of text that need to be synthesized as continuous speech.
The model exposes a configurable guidance parameter \emph{cfg} (with higher values following the conditioning more strictly and lower values allowing more variation), the number of inference time steps (which controls the quality), the temperature, and the maximum number of model generation steps. We experiment with different values using live systems and a subset of LibriSpeech \cite{librispeech} train-100. For final experiments, we choose $cfg=2.0$ a default value of 10 inference steps, 0.9 temperature, and a maximum of 10 model generation steps (which corresponds to 1.6\,s of audio). Putting a rather aggressive cap on model generation steps allowed one to prevent hallucination or generating long silences after the speech. In addition, we restart the continuation mode after silence period, generation of two consecutive phrases or if more than 10 seconds of audio were already generated. Those additional settings are intended to prevent compounding errors that would affect future continuous generations. For all experiments, we used synthesized reference audio using ElevenLabs\footnote{{\url{https://elevenlabs.io/app/speech-synthesis/text-to-speech}}} solution.

\vspace{-3mm}
\section{Evaluation}
\vspace{-3mm}

\subsection{Datasets}
For utility evaluation (WER and UTMOS), we use the development and test splits of LibriSpeech \cite{librispeech}, and for content protection evaluation, we work with SLURP \cite{slurp} and DialogPII \cite{roller2026dialogpii}. SLURP was not designed as a privacy benchmark, but is rare example of NER evaluation benchmark that contains an actual audio, which allows us to perform experiments in a streaming manner and compare them to the results obtained offline and gold transcripts. In our experiments, we used 11514 training and 2974 test transcripts from SLURP and 2913 transcripts from english part of DialogPII.
\subsection{Experiments}
\subsubsection{Latency}
Our first experiment aimed to measure system latency, which is most important parameter for real-life deployments. For this experiment, we chose 100 audio files from the LibriSpeech train-100 and force-aligned their transcripts using WhisperX \cite{whisperX} to obtain gold transcripts for the calculation of ASR latency with continuous Levenshtein alignment \cite{machavcek2025simultaneous}. The same tool was used to calculate utterance-onset end-to-end (E2E) latency, using the emission time of the first audio segment that corresponds to a generated utterance. We also report time to first audio (TTFA) measured from the beginning of the TTS generation and the ratio of durations of generated and input audio. Those measurements do not take into consideration temporal drift during playback caused by differences in duration of input and generated speech. This experiment measures only the ASR + TTS pipeline and does not include the NLP processing time, which will be measured separately for different NLP systems. The experiment was performed on a machine with NVIDIA RTX PRO 6000 Blackwell Max-Q GPU.

\vspace{-3mm}
\subsubsection{Utility}
The utility evaluation of our system mostly follows the metrics proposed by VPC2026~\cite{miao2026voiceprivacy}: 
\textbf{Speech content preservation} is measured by the Word Error Rate (WER) in the anonymized Librispeech dev and test sets~\cite{librispeech}, transcribed by a wav2vec2-based ASR model trained on \textit{LibriSpeech-train-960}. For this evaluation, no PII removal is applied.
\textbf{Emotion preservation} is measured by the Unweighted Average Recall (UAR) on the anonymized IEMOCAP dataset~\cite{busso2008iemocap}, using a wav2vec2-based Speech Emotion Recognition (SER) system, with performance averaged across 5 validation folds.
\textbf{Speech quality} is measured by UTMOS, automatically evaluated by the UTMOSv2~\cite{baba2024t05} system.
\vspace{-3mm}

\subsubsection{Privacy}
The privacy of our system is evaluated upon 2 aspects: the anonymization of the speech components, measured as its capacity to fool a semi-informed ASV system, and the removal of PII, measured as a task of PII detection in text.

\textbf{Speech anonymization:}
The ASV system used is an ECAPA-TDNN with WavLM-large features combined with filterbanks, trained on voice-converted data from the Source Speaker Tracing Challenge (SSTC)~\cite{lislt} which is then fine-tuned on the \textit{Librispeech-train-clean-360}~\cite{librispeech} anonymized data collected during the VPC2024~\cite{tomashenko2024voiceprivacy}.
 The equal error rate (EER) is computed on the gender-mixed trials defined in the VPC2026 from LibriSpeech's test and dev subsets.

\textbf{PII detection:}
For the PII detection task, we focused only on entity types: \emph{person}, \emph{email\_address}, \emph{date} and \emph{time}, \emph{place\_name}, and \emph{relation}, which closely correspond to privacy information, and limit SLURP audio to only headset recordings (resulting in 6281 test recordings). In DialogPII evaluation, we test almost the same subset of entities, except predictions of \emph{date} and \emph{time} are mapped to \emph{datetime} to match the dataset taxonomy. For the transcript-only experiments, each transcript was replicated according to the number of corresponding audio recordings to preserve the same sample distribution as in the audio-based experiments.

We fine-tuned all the PII detection systems using SLURP training transcripts (we use 10\% of the training data for model validation) and simulate streaming by splitting them into blocks of 2 to 4 words. During training, each model sees the left context from previous blocks, but all selections are trimmed to match the text in the current block. Each model is fine-tuned for 10 epochs, and the epoch that corresponds to the highest F1 on the validation subset is selected. The Qwen models were fine-tuned using LoRA \cite{lora}.
For evaluation, adjacent predicted entities of the same type were merged. We evaluated the trained models on gold transcripts and ASR predictions block by block with left context (which corresponds to real system processing). We calculated the micro entity F1 and micro SLU-F1 \cite{slurp}. The first measures exact agreement of detected entities, while the second is designed to assess error propagation in structured E2E-SLU tasks by incorporating word- and character distance based components.

\vspace{-3mm}
\section{Results}
 \begin{table}[t]
    \centering
    \caption{Median of measurements for latency, TTFA and the duration ratio of output and input audio. }
    \label{tab:latency}
    \resizebox{\linewidth}{!}{%
    \begin{tabular}{rrrr}
        \toprule
        \textbf{ASR [ms]} &
        \textbf{TTS (TTFA) [ms]} &
        \textbf{E2E [ms]} &
        \textbf{$T_{out} / T_{in}$} \\
        \midrule
        669 & 41 & 711 & 1.33 \\
        \bottomrule
    \end{tabular}
    }
\end{table}

\vspace{-3mm}

\subsection{Latency}
 The latency results are presented in Table~\ref{tab:latency}. The median ASR latency is consistent with our expectations. Since we use the maximum recognition window of 1.3\,s, and speech can start anytime, on average at least half of that time should contribute to latency. The TTS is extremely fast, although it has to be pointed out that the first frames of audio can contain silence. The difference in the durations of the generated audio is caused by the generation of sentences as separate phrases and the mismatch of the speech rate between the input and voice reference audio.

\begin{table}[t]
\centering
\caption{Utility metrics evaluating linguistic preservation (WER) on LibriSpeech and emotion preservation (UAR) on IEMOCAP, on the original speech.} %
\label{tab:utility_results}
\begin{tabular}{l cc cc}
\toprule

\textbf{System}
& \multicolumn{2}{c}{\textbf{WER} $\downarrow$}
& \multicolumn{2}{c}{\textbf{UAR} $\uparrow$} \\

\cmidrule(lr){2-3}
\cmidrule(lr){4-5}

& \textbf{dev} & \textbf{test} & \textbf{dev} & \textbf{test} \\

\midrule
Orig.                           & 1.80  & 1.84 & 69.08 & 71.06 \\
\midrule
T30\cite{lee2024voice}          & 5.02 & 4.78 & 28.45 & 29.44 \\
T8-4\cite{xinyuan2024hltcoe}    & 3.74 & 3.75 & 30.10 & 30.59 \\
\midrule
{Ours (Streaming)}                          & 11.11 & 11.44 & 29.37 & 29.22 \\
\bottomrule
\end{tabular}%
\end{table}

\begin{table}[ht]
    \centering
                \caption{Privacy results are shown as an EER, evaluated on the Librispeech subsets.}
    \resizebox{\columnwidth}{!}{%
\begin{tabular}{lccccc}
\toprule
\textbf{Split} & \textbf{Orig.}
& \textbf{T30}\cite{lee2024voice}
& \textbf{T8-4}\cite{xinyuan2024hltcoe}
& \textbf{B5}\cite{tomashenko2024voiceprivacy}
& \textbf{Ours} \\
\midrule
\textbf{dev-f} $\uparrow$  & 10.51 & 44.32 & 47.58 & 35.82 & 44.18 \\
\textbf{dev-m} $\uparrow$  & 0.93  & 47.52 & 49.23 & 32.92 & 47.66 \\
\textbf{test-f} $\uparrow$ & 8.76  & 45.07 & 47.44 & 33.95 & 44.37 \\
\textbf{test-m} $\uparrow$ & 0.42  & 44.77 & 48.75 & 34.73 & 45.25 \\
\bottomrule
\end{tabular}
    }
    \label{tab:privacy}
\end{table}

\vspace{-4mm}
\subsection{Utility}
\begin{table*}[t]
\centering
\vspace{-7mm}

\caption{DialogPII and SLURP PII detection results for fine-tuned models.}
\label{tab:PII_results}
\begin{tabular}{l cc cc ccc}
\toprule
& \multicolumn{4}{c}{\textbf{SLURP}} & \multicolumn{3}{c}{\textbf{DialogPII}} \\
\cmidrule(lr){2-5}
\cmidrule(lr){6-8}
\textbf{Model} & \multicolumn{2}{c}{\textbf{Clean transcript}} & \multicolumn{2}{c}{\textbf{ASR chunks}} & \multicolumn{3}{c}{\textbf{Simulated Streaming}} \\
\cmidrule(lr){2-3}
\cmidrule(lr){4-5}
\cmidrule(lr){6-8}
 & Entity F1$\uparrow$ & SLU F1$\uparrow$ & Entity F1$\uparrow$ & SLU F1$\uparrow$ & Entity F1$\uparrow$ & SLU F1$\uparrow$ & Latency [ms]$\downarrow$ \\
\midrule
Qwen 1.5B   & \textbf{87.50} & 89.10 & 51.48 & 66.51 & 56.27 & 64.22 & 46 \\
Qwen 0.5B   & 83.92 & 85.42 &  48.33 & 62.73 & 46.55 & 56.62 & 39 \\
spaCy       & 87.19 & 88.98 & \textbf{54.82} & \textbf{69.01} & 53.27 & 60.60 & \textbf{4} \\
GLiNER Small v2.5 & 82.55 & 87.55 & 50.59 & 66.93 & \textbf{61.22} & \textbf{69.14} & 6 \\
NVIDIA GLiNER PII & 86.89 & \textbf{89.60} & 53.74 & 68.66 & 60.40 & 67.97 & 17 \\
\bottomrule
\end{tabular}
\vspace{-6mm}

\end{table*}

Table \ref{tab:utility_results} shows the utility evaluation results for the proposed system compared to the baselines of the VPC2026.
Emotion preservation shows similar levels to existing non-streaming cascaded anonymization systems, with a UAR at 29.22\%.

However, speech preservation suffers from ASR streaming, showing up to 11.44\% WER on the LibriSpeech test after reconstruction.
The ASR alone achieves a WER of 7.83\%, and the TTS system on ground-truth transcripts yields a WER of 10.38\%, showing that while the first block may be the fastest way to improve PII removal, the TTS also needs improvement for better speech preservation.

Our system also exhibits a 3.21$\pm$ 0.44 UTMOS, compared to the original speech, which is 3.35$\pm$0.28, showing that little to no degradation is added to the quality of the generated speech.
\subsection{Privacy}

 Speech privacy results are shown in Table~\ref{tab:privacy}, which compares our system with the other systems. %
As expected from a cascaded ASR+TTS system, our proposed system outperforms the baseline system, with EER$\geq$44\%, but also places itself in the same range as existing cascaded systems.

The PII detection results are presented in Table~\ref{tab:PII_results}. 
SLURP is a challenging dataset, and our streaming-like ASR system achieves 24\% WER, which contributes to the degradation of PII detection in ASR chunks. The results obtained on DialogPII are shown along with the measured latency in simulated streaming scenarios.
The GLiNER Small v2.5 entity F1 corresponds to 61.55 precision and 59.30 recall, while SLU F1 corresponds to 69.09 precision and 66.88 recall. The results show that simpler NER systems can efficiently replace small LLMs in streaming scenarios, while showing significantly faster processing (4 ms for spaCy vs 46 ms for Qwen 1.5B).
The fine-tuning and evaluation of the NER and LLM models used a non-systematic selection of hyperparameters, mainly default or commonly used values.Therefore, the results presented in Table~\ref{tab:PII_results} should not be interpreted as a careful comparison of their capabilities. The aim of this work is to establish a reasonable lower bound for the robustness of a cascade-like voice privacy system.

\vspace{-3mm}
\section{Conclusion}
\vspace{-3mm}

We presented a streaming-oriented ASR-NLP-TTS cascade that combines speaker anonymization with PII detection for content protection. The Whisper-based configuration achieves an estimated median utterance-onset end-to-end latency of 717\,ms (ASR latency + NER latency + TTFA). In VPC2024 gender-specific trials, EERs range from 44.18\% to 47.66\%, exceeding the baselines evaluated B2-B5 and approaching the results of existing offline cascades. These findings support the feasibility of extending cascaded speech privacy toward streaming operation.

The results highlight the importance of jointly selecting and coordinating cascade components. %
Dedicated NER models offer a promising alternative to small LLMs: on simulated DialogPII streaming, GLiNER Small achieves 61.22\% entity F1 at 6\,ms latency, compared with 56.27\% at 46\,ms for Qwen 1.5B. Thus, efficient content filtering need not require generative language models in the evaluated setting. However, depending on the application, even the simpler dictionary approach could be used to filter PII, and on the other hand, an LLM could perform additional rewriting of the text to hide the information.

Future work will focus on two directions. First, we will reduce end-to-end latency and accumulated delays through adaptive text buffering and coordinated ASR-TTS scheduling with true streaming ASR. Second, the substantial degradation of PII detection on streaming ASR outputs motivates the development of streaming NLP methods that remain robust to recognition errors and incomplete context, enabling more reliable content protection from incremental transcripts.

\subsubsection*{Generative AI Use Disclosure}
Generative AI tools supported software development, debugging, test generation, implementation analysis, LaTeX table generation. They were also used to provide editorial and critical feedback on author-written manuscript text, including its clarity, presentation, and argumentation. All numerical results are derived from the described experiments, with AI-generated scripts used only to process and format outputs. The authors remain fully responsible for the methodology, results, and conclusions.

\bibliographystyle{IEEEbib}
\bibliography{custom}

\end{document}